\documentclass[runningheads]{llncs}
\usepackage[utf8]{inputenc}
\usepackage{mwe} 
\usepackage{graphicx}
\usepackage{float}
\usepackage{hyperref}
\usepackage{tabularx}
\usepackage{multirow}
\usepackage{amsmath}
\usepackage{array}
\usepackage{booktabs}
\usepackage{caption}
\usepackage{subcaption}
\usepackage{multicol}

\title{Towards Field-Ready AI-based Malaria Diagnosis: A Continual Learning Approach}
\titlerunning{Improving Malaria Diagnosis with Continual Learning}

\begin{document}
\author{Louise Guillon\inst{1} \and
Soheib Biga\inst{1} \and
Yendoube E. Kantchire \inst{2} \and
Mouhamadou Lamine Sane \inst{3} \and
Grégoire Pasquier \inst{4} \and
Kossi Yakpa \inst{5} \and
Stéphane E. Sossou \inst{6} \and
Marc Thellier \inst{3,7} \and
Laurent Bonnardot \inst{1} \and
Laurence Lachaud \inst{4} \and
Renaud Piarroux \inst{3,7} \and
Ameyo M. Dorkenoo \inst{2,8}
}
%
\authorrunning{L. Guillon et al.}
%
\institute{MyC, Paris, France \\
\email{lguillon@myc.doctor}\\
\and
CHU Campus, Ministère de la Santé et de l’hygiène Publique, Lomé, Togo \and
Service de Parasitologie-Mycologie, Hôpital de La Pitié-Salpêtrière, Assistance Publique-Hôpitaux de Paris (AP-HP), France \and
CHU of Montpellier, Université de Montpellier, Montpellier, France \and
PNLP Togo, Lomé, Togo \and
CHU Sylvanus Olympio, Lomé, Togo \and
Institut Pierre-Louis d'Epidémiologie et de Santé Publique, INSERM, Sorbonne Université, Paris, France \and 
Faculté des Sciences de la Santé, Université de Lomé, Lomé, Togo
}

\maketitle

\begin{abstract}
Malaria remains a major global health challenge, particularly in low-resource settings where access to expert microscopy may be limited. 
Deep learning-based computer-aided diagnosis (CAD) systems have been developed and demonstrate promising performance on thin blood smear images. 
However, their clinical deployment may be hindered by limited generalization across sites with varying conditions. Yet very few practical solutions have been proposed. 
In this work, we investigate continual learning (CL) as a strategy to enhance the robustness of malaria CAD models to domain shifts. We frame the problem as a domain-incremental learning scenario, where a YOLO-based object detector must adapt to new acquisition sites while retaining performance on previously seen domains. We evaluate four CL strategies, two rehearsal-based and two regularization-based methods, on real-life conditions thanks to a multi-site clinical dataset of thin blood smear images.
Our results suggest that CL, and rehearsal-based methods in particular, can significantly improve performance. These findings highlight the potential of continual learning to support the development of deployable, field-ready CAD tools for malaria.

\end{abstract}


\section{Introduction}

Malaria, a parasitic disease caused by \textit{Plasmodium}, remains a major global health challenge. Despite important efforts for prevention, diagnostic and treatment, malaria accounted for close to 600,000 deaths in 2023 \cite{noauthor_world_nodate}. 
Early and accurate diagnosis is critical to mitigate its impact, with several diagnostic methods available. Rapid Diagnosis Tests (RDT) offer fast and reliable results, but they can suffer from false negatives due to parasite mutations, and false positives can occur after treatment, leading the WHO to recommend microscopic analyses of thick and thin blood smears as the gold standard. Thin smears, in particular, are used to identify \textit{Plasmodium} within red blood cells (RBC).

However, the analysis of thick and thin blood smears requires high levels of expertise, which may be unavailable in remote areas where urgent diagnosis are needed. 
To address this challenge, several works have proposed computer-aided diagnosis (CAD) tools for malaria, based on thin and thick smears \cite{delahunt_automated_2015,guemas_automatic_2024,rajaraman_performance_2019}.
High performance has been achieved at the patient level, with 100\% sensitivity and 95\% specificity for thick smears thanks to the Autoscope \cite{delahunt_automated_2015}. Using smartphone-acquired images, 91.8\%, 92.5\%, 91.1\% were obtained at the patient level for accuracy, sensitivity and specificity respectively \cite{yu_patient-level_2023}. Good performances have also been obtained for thin smears, with accuracies of 98.62\% and 98.44\% reported at the cell and patient levels respectively \cite{liu_aidman_2023}.
Smartphone applications incorporating these algorithms are being developed and tested in the field \cite{nakasi_mobile-aware_2021,rajaraman_performance_2019,yu_patient-level_2023}.

While these works bring massive opportunities, the real-world deployment of such CAD systems remains hindered by a critical challenge: generalization. Indeed, for a CAD system to be used in the field, it must perform well across various clinical environments in spite of deep learning models often sensitive to such variations \cite{zech_variable_2018}.
Malaria CAD is not spared from this challenge, as different clinical sites can exhibit significant domain shifts \cite{delahunt_automated_2015,kabore_addressing_2024,guillon_assessing_2025}. This is particularly critical for routine clinical data which present challenges such as imbalanced dataset and varying image quality due to differences in equipment, staining and operators, making the task even more challenging.
To address the site effect and improve generalization, strategies such as fine-tuning or joint training, can be used \cite{guillon_assessing_2025}. Continual learning (CL) offers another promising approach, where the model is trained incrementally, adapting to new domains or tasks while retaining knowledge from previous ones. CL can be categorized into task, domain, and class incremental scenarios \cite{kumari_continual_2024}. In the context of this article, we are dealing with a domain-incremental scenario where the same task, detecting \textit{Plasmodium}, is performed across different sites. 

Here, we propose the first application of continual learning for malaria CAD. 
We focus specifically on routine thin blood smear images and investigate how CL applied to YOLO models can help overcome generalization, paving the way for more robust, field-ready diagnostic tools.

\section{Related works}

\subsection{Malaria Computer-Aided Diagnosis}
Two main methods have been proposed for thin blood smear images: one-step approaches, which treat the problem as an object detection task, and two-step approaches, which first extract the red blood cells before classifying them.  
In one-step methods, the model is trained to detect parasites on images of a thin smear for example. In this setting, models like Faster RCNN \cite{hung_applying_2017,nakasi_mobile-aware_2021}, the YOLO family \cite{yang_cascading_2020,krishnadas_classification_2022,liu_aidman_2023} and RT-DETR \cite{guemas_automatic_2024} have been used. 
However, the clinical applicability of such malaria CAD systems highly depends on their generalization capabilities across various environments. Although generalization has been largely underexplored in malaria diagnosis, studies have shown that performance can degrade significantly when CAD systems are applied to data from different clinical sites \cite{guillon_assessing_2025,kabore_addressing_2024}, with accuracy dropping by over 20\% at the image level in some cases \cite{guillon_assessing_2025}. These results highlight the need for approaches that can enhance generalization ability of CAD systems, and continual learning could offer a promising solution.

\subsection{Continual Learning}
Continual learning is a deep learning paradigm aiming to address generalization by enabling models to adapt and accumulate knowledge over time, without forgetting previously learned information, commonly known as \textit{catastrophic forgetting}. CL balances between two challenges: plasticity, which allows the model to learn new tasks, and stability, which helps retain knowledge from previous tasks \cite{zhuo_continual_2023}. 
Three main strategies have been introduced: rehearsal, regularization and architectural. The rehearsal strategy involves storing samples from previous tasks in a memory buffer and reusing them during training to prevent forgetting. Regularization-based methods, such as Elastic Weight Consolidation (EWC) \cite{kirkpatrick_overcoming_2017} and Learning Without Forgetting (LWF) \cite{li_learning_2018}, introduce regularization terms in the loss to preserve important parameters and avoid catastrophic forgetting. Architectural methods isolate task-specific parameters.
CL has become an important area of research in medical imaging \cite{kumari_continual_2024}, but to our knowledge, it has not yet been studied in the context of malaria diagnosis and very few studies have implemented CL for YOLO models \cite{monte_teach_2025}.

\section{Methods}

\textbf{Problem Formulation.}\newline
We define the problem as a domain-incremental scenario 
where the goal is to train a model that can effectively adapt to new domains while 
minimizing catastrophic forgetting.
Let $T={T_{1},T_{2},...,T_{T}}$ represent the sequence of tasks.
Each task involves a dataset $D_{t}$ consisting of thin smear images, and the corresponding labels indicating infected RBC.
We seek to train models, $f_{t}$, that learn from the sequence of tasks and must generalize well to each task $T_{t}$.
\newline

\begin{figure}[ht]
    \centering
    \includegraphics[scale=0.27]{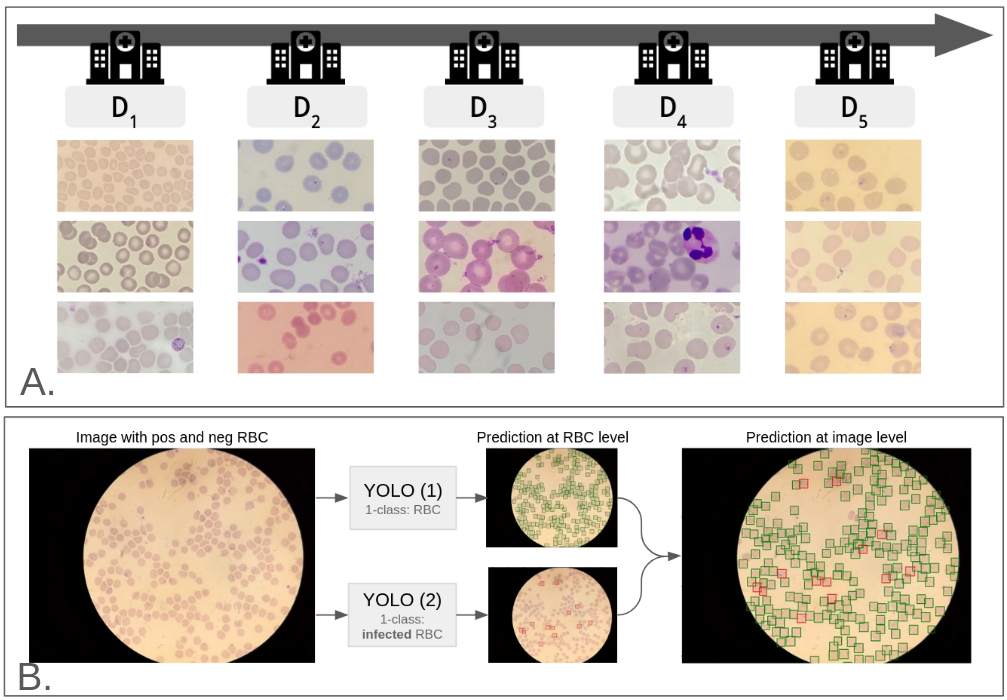}
    \caption{\textit{Workflow overview.} \textbf{A.} Tasks stream. The CL scenario includes fives tasks. Center-cropped images of the 5 sites are presented. \textbf{B}. Malaria diagnosis prediction flow. A field of view image of a thin blood smear is passed to two yolov5 models to detect (1) all RBC and (2) only the infected RBC. The predictions are then merged to obtain the final predictions of negative and positive cells.}
    \label{fig:workflow}
\end{figure}

\noindent \textbf{Datasets.}\newline
We consider five tasks (T=5), each corresponding to a dataset from a distinct site. 
The datasets include the publicly available NIH dataset\footnote{https://data.lhncbc.nlm.nih.gov/public/Malaria/NIH-NLM-ThinBloodSmearsPf/index.html}, captured in Bangladesh \cite{kassim_clustering-based_2021}, and four in-house datasets collected from four sites in two countries, in endemic and non-endemic areas.
The datasets were acquired through a smartphone connected to a microscope at x1000 magnification during clinical routine.
Since our in-house datasets are field clinical data, they present varying distributions of positive and negative images, as well as diverse \textit{Plamosdium} species. These factors complicate generalization, making the task more challenging \cite{kabore_addressing_2024}.  
Table \ref{tab:data_description} presents the datasets and Fig. \ref{fig:workflow} A. shows cropped zoomed-in examples.
\newline


\begin{table}
    \centering
    \caption{\textbf{Description of the datasets.} Each dataset $D_{t}$ corresponds to a site. "positive images" refers to images that contain at least one parasite.}
    \begin{tabularx}{\textwidth}{|c|>{\centering\arraybackslash}X|>{\centering\arraybackslash}X|>{\centering\arraybackslash}X|>{\centering\arraybackslash}X|>{\centering\arraybackslash}X|>{\centering\arraybackslash}X|>{\centering\arraybackslash}X|>{\centering\arraybackslash}X|>{\centering\arraybackslash}X|}
        \hline
        & $D_{1}$ & $D_{2}$ & $D_{3}$ & $D_{4}$ & $D_{5}$ \\
       \hline
       \multicolumn{6}{|c|}{\textbf{Train}}  \\
       \hline
        \textbf{\# patients} & 92 & 155 & 40 & 21 & 22  \\
       \hline
       \textbf{\# images} & 1316 & 775 & 160 & 95 & 151  \\
       \hline
       \textbf{\# positive images} & 1087 & 488 & 70 & 48  & 16\\
       \hline
       \multicolumn{6}{|c|}{\textbf{Test}}  \\
       \hline
        \textbf{\# patients} & 28 & 38 & 10 & 4  & 5 \\
       \hline
       \textbf{\# images} & 323 & 190 & 40 & 26 & 32 \\
       \hline
       \textbf{\# positive images} & 237 & 124 & 28 & 10  & 6 \\
       \hline
    \end{tabularx}
    \label{tab:data_description}
\end{table}

\noindent \textbf{Malaria Prediction.} \newline
To predict malaria, we seek to detect infected RBC. We tackle this problem as an object detection task where we detect parasites thanks to the YOLO framework \cite{redmon_you_2016,jocher_ultralyticsyolov5_2022} which has proved to be effective in this context \cite{dave_codamal_2024,guemas_automatic_2024,krishnadas_classification_2022,yang_cascading_2020}.
Following the method of \cite{guillon_assessing_2025}, we train two yolo models. One is trained to detect all RBC and the second to detect infected RBC. Detections are then merged (see Fig. \ref{fig:workflow}. B.) 
An image is considered positive if at least one RBC is predicted infected. 
\newline

\noindent\textbf{Continual Learning Strategies}

\noindent \textbf{Reference models.}
\begin{itemize}
    \item \textbf{Baseline}: to underline the need to address generalization, the first strategy consists in the model trained on $T_{1}$, applied to the following tasks, $T_{2}$,...,$T_{T}$. We consider this to be the lower bound since no adaptation is done. 
    \item \textbf{Incremental Joint Training}: the model is trained incrementally using all available data from the previous tasks. 
    It represents the best case scenario.
\end{itemize}

\noindent \textbf{Regularization strategy.}
\begin{itemize}
    \item 
    \textbf{EWC} \cite{kirkpatrick_overcoming_2017} adds a regularization term to the loss function that penalizes large changes in the model parameters that are crucial for previously learned tasks. 

    \item \textbf{LWF} \cite{li_learning_2018} leverages knowledge distillation to prevent forgetting. It encourages the model to preserve the feature maps and predictions from prior tasks.
\end{itemize}

\noindent \textbf{Rehearsal strategy.} For both methods, the memory buffer is composed of 80\% positive images to ensure 
adequate representation of positive cases, as each positive image contains many more negative RBC than positive ones. 
The buffer is capped at 125 images per site and, in case of sites with fewer images, the memory buffer size is reduced to no more than 50\% of $D_{t}$.
\begin{itemize}
\item \textbf{Naive Replay}: examples from previous tasks are randomly chosen.

\item \textbf{Confidence Replay}: inspired by UACL \cite{sadafi_continual_2024} which leverages the model's confidence to select the samples, examples are selected according to YOLO confidence. Specifically, the examples which contain detections with the lowest confidence are selected.
\end{itemize}


\noindent \textbf{Experimental Design.} 

\noindent \textbf{Training strategy:} 
For all trainings, we performed a 3-fold cross validation, ensuring that all images from a single patient were either in the training, validation, or test set to prevent data leakage. Models were trained for 50 epochs, applying early stopping. 
For LWF and EWC, we choosed the regularization term $\lambda$ via gridsearch: for EWC, $\lambda$=[0.5, 10, 1000], and for LWF, $\lambda$=[0.5, 1, 10]. The results were overall similar, finally $\lambda$=10 was chosen for EWC and $\lambda$=1 for LWF.

\noindent \textbf{Evaluation:} Sensitivity and specificity, alongside accuracy, are crucial for malaria CAD. Metrics are reported at both the RBC and image levels. We assume access to a test set for each of the $T$ tasks, enabling the computation of the global train-test performance matrix $P \in P^{T \times T}$, where $P_{t, i}$ is the performance of model $f_{t}$ on task $i$'s test data \cite{kumari_continual_2024,lopez-paz_gradient_2017}. From this, we compute the average performance of $f_{T}$ on all test sets, providing a global measure of generalization across sites (eq.\ref{av_acc}). Specific to continual learning, we also evaluate backward transfer (BWT) and forward transfer (FWT) of accuracy (eq. \ref{bwt}, \ref{fwt}) \cite{lopez-paz_gradient_2017}, which reflect the model's ability to retain knowledge (BWT) and generalize to new tasks (FWT). In eq. \ref{fwt}, $\bar{b}_{t}$ represents the performance of a model with random weights.

\begin{equation}
    Av. \space perf = \frac{1}{T}\sum_{i=1}^{T} P_{T,i}
    \label{av_acc}
\end{equation} 

\begin{multicols}{2}
\begin{equation}
    BWT = \frac{1}{T-1}\sum_{i=1}^{T-1} (P_{T,i}-P_{i,i})
    \label{bwt}
\end{equation} 

\begin{equation}
    FWT = \frac{1}{T-1}\sum_{i=2}^{T} (P_{i-1,i}-\bar{b}_{i})
    \label{fwt}
\end{equation}
\end{multicols}


\section{Experiments and Results}

\begin{figure}
    \centering
    \includegraphics[scale=0.19]{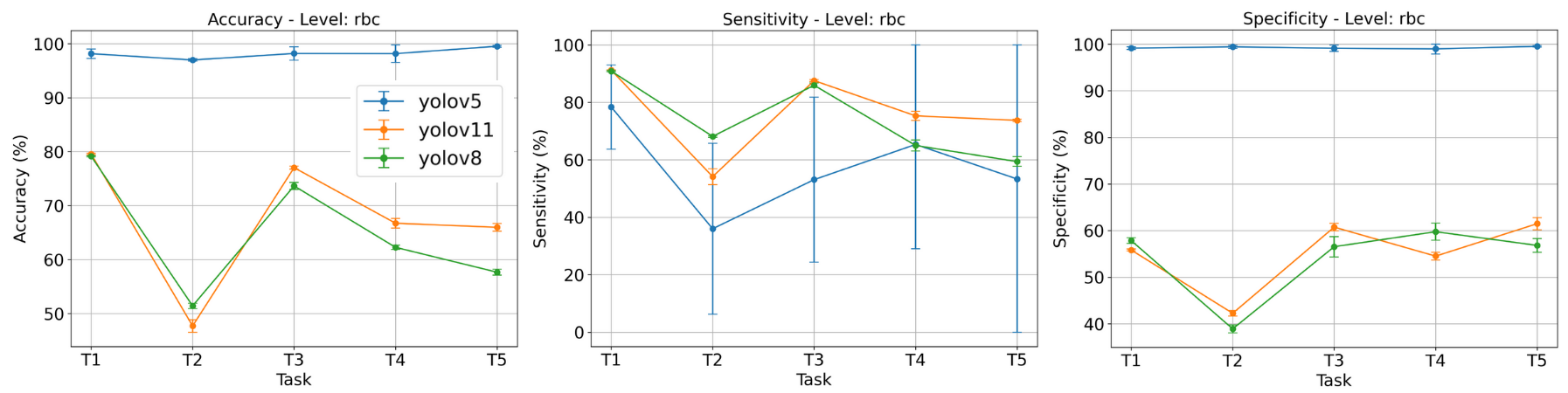}
    \caption{\textit{Performance during the tasks stream for YOLOv5, YOLOv8 and YOLOv11.} Each value represents $P_{1,i}$ for either accuracy, sensitivity, or specificity at RBC level. The error bars correspond to the standard deviation over the three folds.}
    \label{fig:v811}
\end{figure}

We trained our baseline model with Yolov5s, Yolov8s and Yolov11s which gave significantly better results for Yolov5s, as v8 and v11 led to poor generalization (Fig. \ref{fig:v811}). Therefore, we applied the continual learning strategies only to Yolov5.  

The final performance (Tab. \ref{tab:average_perf}) presents the model's ability to generalize across multiple sites, averaging the accuracy, sensitivity, and specificity after training on all tasks. Among the methods tested, Replay conf stands out with the highest accuracy (98.53\%) and specificity (99.26\%) at the RBC level, and the best trade-off at the image level. Indeed, in spite of a weaker sensitivity, Replay conf seems to be the most promising  approach, even outperforming the upper bound at the image level. 
It is also interesting to note that Replay conf method seems to be the most robust one, with less variability. Conversely, regularization-based methods perform significantly worse at both levels. 
Finally, Replay conf is also the methods achieving the best backward transfer (0.018), suggesting that instead of catastrophic forgetting, the model becomes better along the tasks,  while having a good forward transfer.

\begin{figure}
    \centering
    \includegraphics[scale=0.34]{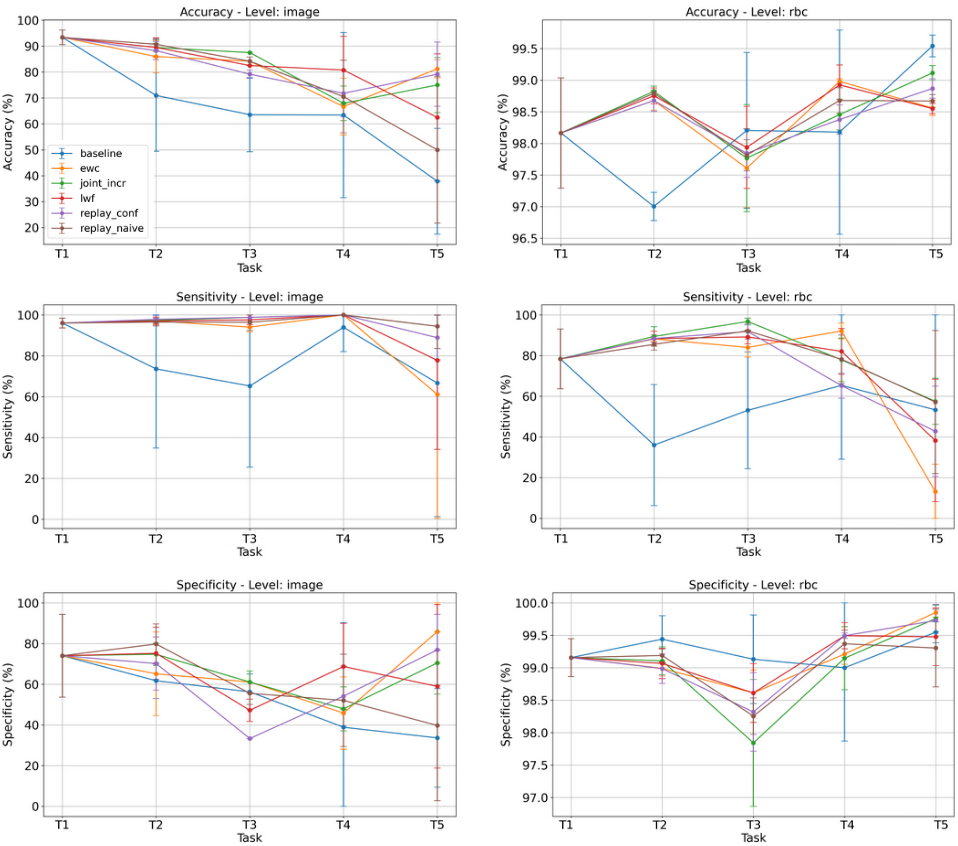}
    \caption{\textit{Performance during the tasks stream.} Each value represents $P_{i,i}$ for either accuracy, sensitivity, or specificity at image and RBC levels, except for baseline which is $P_{1,i}$. The error bars correspond to the standard deviation over the three folds.}
    \label{fig:curves}
\end{figure}

Fig. \ref{fig:curves} shows the evolution of performance during the training process. 
First, we note that, as expected, baseline has difficulties to generalize; sensitivity, both at image and RBC levels is particularly impacted, dropping from near 80\% to less than 40\% from task 1 to 2 at RBC level. 
For all CL methods, at image level, the accuracy of the different approaches slowly decreases but remains better than baseline. Indeed, at RBC and image levels, CL methods significantly improve the sensitivity, except for the last task. This could be due to the image sampling which included far fewer positive images in $T_{5}$.
Overall, Replay conf seems to maintain a rather steady performance throughout the tasks.
Similar to the results in Tab. \ref{tab:average_perf}, most CL methods, at each step, enable to reduce the variability thus increasing robustness.
Lastly, it is interesting to note that for the third task, specificity particularly drops. This could be due to the fact that some negative images demonstrate artifacts such as stains which can be wrongly predicted as parasites.
All in all, based on the curves and the tables, confidence-based replay seems to be the most promising strategy with a balanced performance.

\begin{table}[ht]
\centering
\caption{Average performance across models and metrics (mean ± std). Best values are indicated in bold. BWT and FWT are computed based on the accuracy.}
\label{tab:average_perf}
\setlength{\tabcolsep}{4pt}
\subcaption{RBC level}

\begin{tabular}{lcccccc}
\toprule
\textbf{Approach} & \textbf{Av. Acc} & \textbf{Av. Sens} & \textbf{Av. Spe} & \textbf{BWT} & \textbf{FWT} \\
\midrule
\textbf{Baseline} & 98.22  $\pm $  \scriptsize{0.21} & $57.24  \pm $  \scriptsize{15.00} & $99.26  \pm $  \scriptsize{0.20} & - & - \\
\hline
\textbf{EWC} & $97.09  \pm $  \scriptsize{0.71} & $24.20  \pm $  \scriptsize{19.78} & $99.20  \pm $  \scriptsize{0.26} & -0.16 & 0.20 \\
\textbf{LWF} & $97.91  \pm $  \scriptsize{0.48} & $54.76  \pm $  \scriptsize{21.88} & $99.20  \pm $  \scriptsize{0.31} &-0.069 & 0.16 \\
\textbf{Replay naive} & $98.38  \pm $  \scriptsize{0.17} & $\textbf{80.94}  \pm $  \scriptsize{8.22} & $98.83  \pm $  \scriptsize{0.33} & -0.0057 & \textbf{0.23} \\
 \textbf{Replay conf} & $\textbf{98.53} \pm $ \scriptsize{0.19} & $72.02 \pm $ \scriptsize{8.26} & $\textbf{99.26} \pm $ \scriptsize{0.03} & \textbf{0.018} & 0.21 \\ 
\hline
\textbf{Joint incr} & $98.79  \pm $  \scriptsize{0.05} & $84.32  \pm $  \scriptsize{1.59} & $99.20  \pm $  \scriptsize{0.03} & - & - \\
\bottomrule
\end{tabular}

\vspace{3mm}
\subcaption{Image level}
\begin{tabular}{lcccccc}
\toprule
 \textbf{Approach} &  \textbf{Av. Acc} &  \textbf{Av. Sens} &  \textbf{Av. Spe} &  \textbf{BWT} &  \textbf{FWT} \\
\midrule
\textbf{Baseline} & $65.84  \pm $  \scriptsize{9.65} & $79.10  \pm $  \scriptsize{15.54} & $52.92  \pm $  \scriptsize{12.87} & - & - \\
  \hline
 \textbf{EWC} & $70.00  \pm $  \scriptsize{12.69} & $53.42  \pm $  \scriptsize{33.63} & $\textbf{91.19}  \pm $  \scriptsize{8.67} &-1.54 & \textbf{3.52} \\

 \textbf{LWF} & $75.76  \pm $  \scriptsize{4.90} & $86.71  \pm $  \scriptsize{13.59} & $63.72  \pm $  \scriptsize{20.73} & -0.75 & 3.48 \\

 \textbf{Replay naive} & $74.50  \pm $  \scriptsize{5.97} & $\textbf{97.87}  \pm $  \scriptsize{2.31} & $50.19  \pm $  \scriptsize{11.37} & -0.41 & 3.40 \\
 \textbf{Replay conf} & $\textbf{81.95} \pm $ \scriptsize{1.05} & $94.97 \pm $ \scriptsize{2.21} & $64.74 \pm $ \scriptsize{3.41} & \textbf{-0.050} & 3.34 \\
\hline
 \textbf{Joint incr} & $78.51  \pm $  \scriptsize{1.56} & $97.21  \pm $  \scriptsize{1.81} & $55.54  \pm $  \scriptsize{3.98} & - & -\\
\bottomrule
\end{tabular}
\end{table}

\section{Conclusion}

In this work we compared four continual learning strategies for malaria computer-aided-diagnosis based on YOLO.
We confirmed that generalization remains a significant challenge, particularly in terms of sensitivity. Our results suggest that continual learning strategies can help address this challenge and perform similarly or better than incremental joint training, which is considered as the upper bound. In particular, rehearsal-based approaches seem to be the most promising and leveraging the confidence score output by YOLO avoids catastrophic forgetting. 
A previous work had demonstrated that incorporating 200 random samples from new sites could significantly improve generalization \cite{guillon_assessing_2025}. Here, our results suggest that even with fewer samples, selecting them  based on YOLO's confidence rather than randomly, enhances performance, particularly at the image level.
Given the clinical focus of this work, it relies on routine data that can be constrained in size and variable in quality. Thus, we consider this is a preliminary step toward broader real-world implementation.
Future works should focus on expanded data acquisition, with particular attention to \textit{Plasmodium} species, as well as further algorithmic investigations.
Key areas include assessing the impact of task order, especially as data distribution varies, and examining the potential of foundation models. For instance, DinoBloom, a model proposed for white blood cells classification \cite{koch_dinobloom_2024} could be adapted for this task.
\newline
\textbf{Prospect of Application.} 
This research will be applied in the context of our partnership between a start-up, two hospitals in non-endemic areas, two reference training centers and a dozen of health centers  in Africa in order to improve diagnosis thanks to training and quality assessment. The algorithm will be embedded in a smartphone application which can be directly used on the field.

\subsubsection{\ackname}
The authors deeply thank Quentin Dubois, Hugo Riou, Adel Khiter and Alexandre Merle from the MyC team and Moussa Amadou, Emmanuel T. Doumongue and Kokou Mawusé Etse for their involvement in the project.
\subsubsection{Disclosure of Interests.}
The authors have no competing interests to declare that are
relevant to the content of this article.

\bibliographystyle{splncs04}
\bibliography{bibliography}

\end{document}